\documentclass{aa}  

\usepackage{graphicx}
\usepackage{txfonts}
\usepackage{comment}
\PassOptionsToPackage{hyphens}{url}
\usepackage[colorlinks=true,allcolors=blue]{hyperref}
\usepackage[dvipsnames]{xcolor}
\usepackage{enumitem}

\newcommand{\erosita}{\textit{eROSITA}}
\newcommand{\chandra}{\textit{Chandra}}
\newcommand{\xmm}{\textit{XMM-Newton}}

\newcommand{\swift}{\textit{Swift}}

\newcommand{\hlxone}{ESO~243-49~HLX-1}

\newcommand{\ergcms}{erg\,cm$^{-2}$\,s$^{-1}$}
\newcommand{\ergs}{erg\,s$^{-1}$}

\newcommand{\candone}{4XMM J231818.7-422237}
\newcommand{\candtwo}{4XMM J161534.3+192707}
\newcommand{\candthree}{4XMM J160553.3+324734}
\newcommand{\candfour}{4XMM J161604.0-223726}

\begin{document}

   \title{The population of hyperluminous X-ray sources as seen by \xmm{} }
    \titlerunning{The population of HLXs as seen by \xmm{}}
   \subtitle{}

   \author{Roberta Amato
          \inst{1,2}
          \and Erwan Quintin \inst{3}
          \and Hugo Tranin \inst{4,5,6} 
          \and Andr\'es G\'urpide \inst{7}
          \and Natalie Webb \inst{2}
          \and Olivier Godet \inst{2}
          \and Gian Luca Israel \inst{1}
          \and Matteo Imbrogno \inst{1}
          \and Elias Kammoun \inst{8}
          \and Maitrayee Gupta \inst{9}
          }

   \institute{INAF -- Osservatorio Astronomico di Roma, Via Frascati 33, I-00078, Monte Porzio Catone (RM), Italy \\ \email{roberta.amato@inaf.it} 
   \and  IRAP, CNRS, Université de Toulouse, CNES, 9 Avenue du Colonel Roche, 31028 Toulouse, France
   \and European Space Agency (ESA), European Space Astronomy Centre (ESAC), Camino Bajo del Castillo s/n, 28692 Villanueva de
la Cañada, Madrid, Spain 
\and Institut de Ciències del Cosmos (ICCUB), Universitat de Barcelona (UB), c. Martí i Franquès, 1, 08028, Barcelona, Spain 
\and Departament de Física Quàntica i Astrofísica (FQA), Universitat de Barcelona (UB), c. Martí i Franquès, 1, 08028, Barcelona,
Spain 
\and Institut d’Estudis Espacials de Catalunya (IEEC), c. Gran Capità, 2-4, 08034, Barcelona, Spain
\and
School of Physics \& Astronomy, University of Southampton, Southampton, Southampton SO17 1BJ, UK
\and Cahill Center for Astrophysics, California Institute of Technology, 1216 East California Boulevard, Pasadena, CA 91125
\and 
Astronomical Institute of the Czech Academy of Sciences, Boční II 1401/1, 14100 Praha 4, Czech Republic
    }           

   \date{}

\abstract{Ultraluminous and hyperluminous X-ray sources (ULXs and HLXs) are among the brightest astrophysical objects in the X-ray sky. While ULXs most likely host stellar-mass compact objects accreting at super-Eddington rates, HLXs are compelling candidates for accreting intermediate-mass black holes.
However, HLXs are predominantly found in distant galaxies ($d\gtrsim100$\,Mpc), where the chances of source confusion and misidentification with active galactic nuclei (AGNs) or other transient X-ray phenomena are high.}
{Our goal is to produce a clean sample of HLXs, by removing possible contaminants, and characterise the spectral properties of the remaining population. This sample can then be used to determine whether different types of astrophysical objects coexist within the same class.}
{Starting with a set of 115 HLXs detected by \xmm, we identified and removed contaminants (AGNs, X-ray diffuse emission detected as point-like, and tidal disruption event candidates) and retrieved 40 sources for which \xmm\ spectra are available. We fitted them with an absorbed power law model and determined their unabsorbed luminosities and hardness ratios. We constructed the hardness-luminosity diagram, compared the results with the spectral properties of the HLX prototype, \hlxone{}, and conducted a deeper analysis on a few promising candidates.}
{The resulting HLX population spans a luminosity range from $1\times10^{41}$\,\ergs{} to nearly $10^{43}$\,\ergs\ and is homogeneously spread in hardness between 0.5 and 5. Half of the population has hardness ratios higher than a typical AGN, and could be considered the extension of the ULX population at higher energies. We found four very soft outliers, which are characterised by steep power law spectra and no X-ray emission above 1--2\,keV, similarly to ESO~243-49~HLX-1. Those with multi-epoch archival data show changes in luminosity up to almost two orders of magnitudes. }  
{We show that sources currently identified as HLXs can be more diverse than ULXs and disentangling between different types of objects is not trivial  with currently available data. New observations would be beneficial to expand the current sample and uncover the true nature of many objects of this class, some of which show very similar characteristics to \hlxone.} 

             
   \maketitle

%

\section{Introduction}
\label{sec:intro}

Ultraluminous X-ray sources (ULXs) are defined as point-like, X-ray sources with luminosities exceeding 10$^{39}$\,\ergs{}, found in the off-nuclear regions of nearby galaxies \cite[see reviews][]{Kaaret2017, King2023_review,PintoWalton2023_review,Fabrika2021_review}. This luminosity threshold corresponds to the Eddington limit of a 10\,M$_\odot$ accreting black hole (BH), under the assumption of isotropic emission. 
ULXs can host either neutron stars (NSs) or BHs, accreting matter from a companion star at super-Eddington mass transfer rates. For a few ULXs the compact object is unequivocally a NS, as they show pulsations of the order of seconds \citep[e.g.][]{Bachetti2014,Israel2017_Science,Israel2017b,Carpano2018,Fuerst2016,RodriguezCastillo2020,Sathyaprakash2019} and are commonly referred to as pulsating ULXs (PULXs). For the remaining (and the great majority of) sources, which do not show pulsations, it is not possible to determine with certainty the nature of the compact object. Nevertheless, spectral studies suggest that the hardest ULXs are more likely to host NSs, while the softest would be more consistent with hosting stellar-mass BHs \citep{Gurpide2021a_ULXsample,Pintore2017,ReyeroSerantes2024}.

It is not clear whether the most luminous sources, those with X-ray luminosities $\geq10^{41}$\,\ergs{}, commonly referred to as hyperluminous X-ray sources \citep[HLXs,][]{Matsumoto2003,Gao2003}, are just an extension of the ULX class at higher luminosities, represent heavier accreting objects, such as intermediate-mass BHs (IMBHs, with masses $10^2$--$10^5$\,M$_\odot$; see e.g. \citealt{ColbertMushotzky1999}), or are a collection of different types of sources altogether. This doubt is strengthened by the presence of two different objects in the HLX class. The first is the PULX NGC 5907 ULX-1, which reaches a peak luminosity of $2\times10^{41}$\,\ergs{} \citep{Israel2017_Science}. The second is the IMBH candidate \hlxone{} \citep{Farrell2009}, which reaches a peak luminosity of $2\times10^{42}$\,\ergs{} and has a mass estimate for the BH of $9\times10^3$--$9\times10^4$\,M$_\odot$ \citep{Webb2012}.
While NGC 5907 ULX-1 shows characteristics that are similar to other (P)ULXs \citep[e.g][]{Gurpide2021a_ULXsample}, \hlxone{} is more similar to BH binaries rather than ULXs \citep{Servillat2011,Soria2021,Godet2009,Godet2014,Straub2014}. 
This suggests that the HLX population must be made up of at least two different classes of objects, whose prototype could be \hlxone{} and NGC 5907 ULX-1, respectively \citep[see e.g. the recent work of ][]{MacKenzie2023}. 

ULXs and HLXs constitute the high-luminosity end of the X-ray luminosity function (XLF) of X-ray binaries in other galaxies, which has a power-law shape, with a break at $L_\mathrm{X}\sim10^{40}$\,\ergs{} \citep{Tranin2024_cat,Mineo2012}. Whether this break is real, with HLXs being the tail of the ULX XLF, or whether it is due to observational biases is a matter of debate. Additionally, unlike the ULXs, HLXs seem to be equally distributed in elliptical and spiral galaxies \citep{Tranin2024_cat}. While ULXs are typically found in galaxies within a few hundred Mpc (with many even within the first 50\,Mpc; see e.g. the ULX catalogues \citealt{Kovlakas2020_ULXcat, Walton2022_ULXcat, Tranin2024_cat}), HLXs are significantly more distant, with very few sources located within 100\,Mpc. Although this could be partly due to observational biases (as closer ULXs are easier to detect) and sample incompleteness, it is clear that the HLX population predominately lies beyond 100\,Mpc. As a consequence, while one can obtain good-quality data for ULXs with just a few tens of ks of observations with current X-ray facilities, much longer exposures are needed for HLXs (or for ULXs at very high distances, $d>100$\,Mpc), in spite of their higher intrinsic luminosities. Hence, HLXs remain overall less well studied than ULXs, with the exception of the aforementioned ESO 243--49 HLX-1, which has been repeatedly observed by \xmm{} and \chandra{} and still is extensively monitored by \swift{}  \citep[e.g.][]{Webb2023_TDEQPE, Lin2020_HLX1monitor}.

The aim of this paper is to take a deeper look at the HLX population to investigate their nature, on the basis of their luminosity and spectral hardness, to see if different populations can be identified. We collected a sample of 115 HLXs and HLX candidates observed by \xmm\ from the most recent catalogue of ULXs and HLXs of \citet{Tranin2024_cat} (Sect.\,\ref{sec:SampleSelection}), identified and excluded several contaminants and fitted the spectra of the remaining sources to build the hardness-luminosity diagram (HLD, Sect.\,\ref{sec:SpectralAnalysis}). We carried out a more detailed analysis on \hlxone{} (Sect. \,\ref{sec:SpectralAnalysis}) 
and a on few selected promising sources (Sect.\,\ref{sec:AnalysisSelectedSources}). Our conclusions are presented in Sect.\,\ref{sec:Conclusions}.

\section{Sample selection}
\label{sec:SampleSelection}

We considered the HLX sample presented in the catalogue of \citet{Tranin2024_cat}, which collects ULXs and HLXs from the three main X-ray catalogues, namely \chandra{} CSC2 \citep{Evans2010_ChandraCat, Evans2019_ChandraCat}, \swift/XRT 2SXPS \citep{Evans2020_SwiftCat}, and \xmm{} 4XMM-DR12 \citep{Webb2020_XMMCat}. In the catalogue, the association of the sources with their host galaxies was made through cross-correlation with the compilation of galaxy catalogues GLADE \citep{Dalya2018_GLADE}. 
According to the significance of their association, the authors divided the sources into three categories: ``robust'', ``weak'', and ``galaxy pair'' candidates. Robust sources have no detected counterpart in other wavelengths, reducing their chances to be background active galactic nuclei (AGN) or dwarf satellite galaxies, while both weak and galaxy pairs candidates have association with a detected optical counterparts \citep[for more details see][]{Tranin2024_cat}.  

The automated probabilistic method developed by \citet{Tranin2022} guarantees a contamination rate $\leq$2\% of the ULX sample alone. However, for the HLXs this method can produce significant biases, with sources misclassified as stars or AGN when associated with strong optical counterparts. Hence, the authors applied the following further selection criteria for a more secure HLX identification: (1) a mean observed X-ray luminosity $\geq10^{41}$ erg s$^{-1}$ and a signal-to-noise ratio $S/N\geq3$; (2) no evident contaminant from visual inspections; (3) when an optical counterpart is present (weak sources and galaxy pairs), the redshifts between the host galaxy and the counterpart must be consistent (at 1$\sigma$). 
The catalogue does not include information on the peak luminosity of each source (intra-observational or long-term), nor upper limits on missed detections. Our sample is therefore made of sources with average HLX luminosities determined over the course of a single observation. Less luminous X-ray sources undergoing sporadic hyperluminous outbursts might be excluded .

The collection consists of 191 sources, 11 of which are observed in all three X-ray catalogues. In this sample, the contamination rate among robust and weak HLX candidates is expected to be of $\sim$30\% \citep{Tranin2024_cat}. With the aim of carrying out a systematic spectral analysis, we decided to use only data from the 4XMM-DR12 catalogue\footnote{\url{http://xmm-catalog.irap.omp.eu/}}, which contributes for the majority of the sources in the HLX sample (115, 
against 75 for CSC2 and 13 for 2SXPS).

\section{Spectral analysis}
\label{sec:SpectralAnalysis}

The spectral properties of the HLXs and HLX candidates reported in the X-ray catalogues are obtained from spectral fits with simple models (power laws or black bodies, not suited to model ULX-like spectra) and are not corrected for the Galactic and local absorptions. This is the case of the 4XMM catalogue, which  estimates fluxes in different energy bands using an absorbed power law with fixed values of the equivalent hydrogen column $N_\mathrm{H}=3\times10^{20}$\,cm$^{-2}$ and the photon index $\Gamma=1.7$ for all sources \citep{Webb2020_XMMCat}. Therefore, we decided to carry out a more refined spectral analysis of the data to obtain more accurate flux measurements and hardness ratios (HR), as this approach has proven fruitful to categorise and investigate the nature of ULXs \citep[e.g.][]{Gurpide2021a_ULXsample}.

First, we retrieved all available source spectra for all three EPIC cameras on board \xmm{} \citep[pn, MOS1 and MOS2,][]{Struder2001_pn,Turner2001_mos} from the 4XMM-DR12 archive. The \xmm{} Pipeline Processing System (PPS)\footnote{\url{https://www.cosmos.esa.int/web/xmm-newton/pipeline}.} only generates spectra for sources with $\geq100$ counts. These are then rebinned to a minimum of 5 bins ($\geq$20 counts per bin). 
Out of the 115 \xmm{} HLX candidates, we retrieved the spectra of 64 sources, together with their corresponding background spectra and ancillary files (arfs), while the canned response matrices files (rmfs, update 2021-12-09) were downloaded from the \xmm{} online repository\footnote{See \url{https://www.cosmos.esa.int/web/xmm-newton/epic-response-files}.}. 

Before performing the spectral fits, we inspected the candidates labelled as ``galaxy pairs'' to verify their classification. We found that 13 of them are actually confirmed or candidate AGN, positionally coincident with a galaxy, often in pairs or clusters of galaxies. 
The remaining 3 sources are very faint and their spectra are dominated by the background so we decided to exclude them from the current sample. We were left with 48 sources. 

We simultaneously fitted all three EPIC instrument spectra within the python interface of Xspec \citep{Arnaud1996}, PyXspec\footnote{\url{https://heasarc.gsfc.nasa.gov/xanadu/xspec/python/html/index.html}.} (version 1.2.2, with Xspec version 12.10.1), with abundances from \citet{Wilms2000} and cross-sections from \citet{Verner1996}. All the spectra were fitted in the 0.3--10\,keV energy range and the goodness of the fits was evaluated with a chi-squared statistics. We used the same absorbed power law model (\texttt{tbabs$\times$pow}) as in the 4XMM catalogue, but leaving $N_\mathrm{H}$, $\Gamma$, and the power law normalisation free to vary. When the best-fit value of the absorption column was below the Galactic hydrogen column density along the line of sight \citep{HI4PICollaboration2016}, as computed through the python package \texttt{gdpyc} \citep{gdpyc_ruiz_2018}, we fixed it to the average HI4PI value. In total, 32 spectra had a free $N_\mathrm{H}$ and 16 fixed. We rejected all the fits with $\chi^2_\mathrm{red}>2$ and/or null hypothesis probability (n.h.p.) $<0.01$ (cf. Appendix\,\ref{sec:ExcludedSources}) and the background-dominated ones (ObsIDs. 0691590101a and 0502690601). Among the rejected sources, one is ESO 243–49 HLX-1 (cf. Sect.\,\ref{sec:HLX1}), and another is a tidal disruption event (TDE) candidate \citep[4XMM J215022.4-055109,][]{Lin2018}. We refined two further fits with bad statistics either by freezing the $N_\mathrm{H}$ to the HI4PI value (ObsIDs. 0201901801 and 0110870101) and adding a cross-calibration constant between the pn and MOS data (ObsID.0110870101 as well). At the end, we obtained 40 statistically acceptable spectral fits, one for each source.

\begin{figure*}
    \resizebox{\hsize}{!}
    {\includegraphics[]{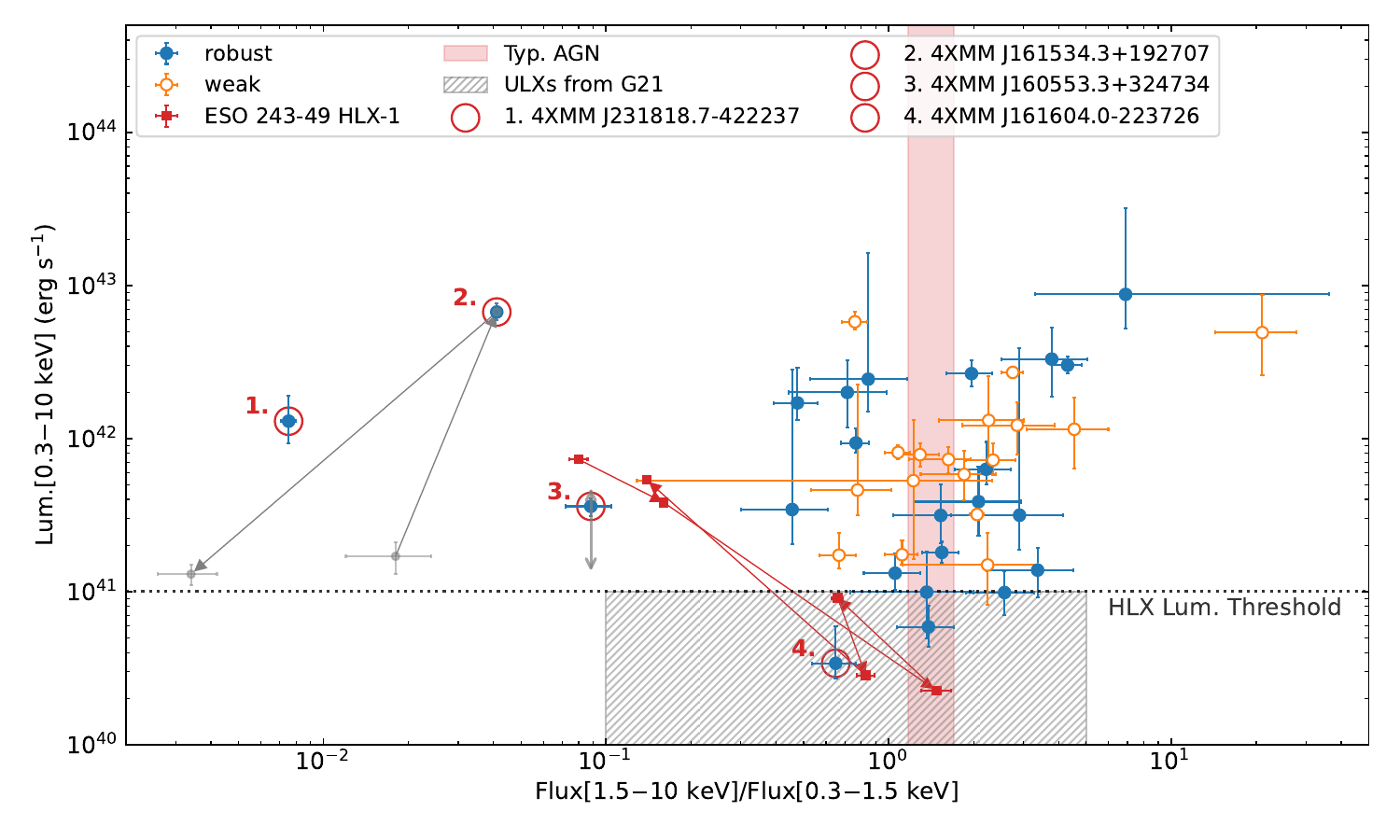}}
    \caption{Hardness-luminosity diagram (HLD) of the sample of HLX candidates considered in this work. Blue filled dots are for sources with a robust association to the host galaxy and empty orange dots for weak associations. The red vertical area indicates the position on the HLD of a typical AGN, with a power law index between 1.8 and 2 (fluxes computed with WebPIMMS, see \url{https://heasarc.gsfc.nasa.gov/cgi-bin/Tools/w3pimms/w3pimms.pl.}). The grey dashed area correspond to the parameter space covered in \citet{Gurpide2021a_ULXsample}. The dotted line indicates the conventional HLX luminosity threshold of $1\times10^{41}$\,\ergs.The long-term spectral evolution of ESO 243--49 HLX-1 is shown in red squares and arrows. The four most interesting candidates analysed in this work (cf. Sect.\,\ref{sec:AnalysisSelectedSources}) are marked with red circles and numbered; for those with available data the long-term evolution or the HR range is reported with grey data points and arrows. }
    \label{fig:HLD}
\end{figure*}

\begin{figure}
    \centering
    \resizebox{.75\hsize}{!}
    {\includegraphics[]{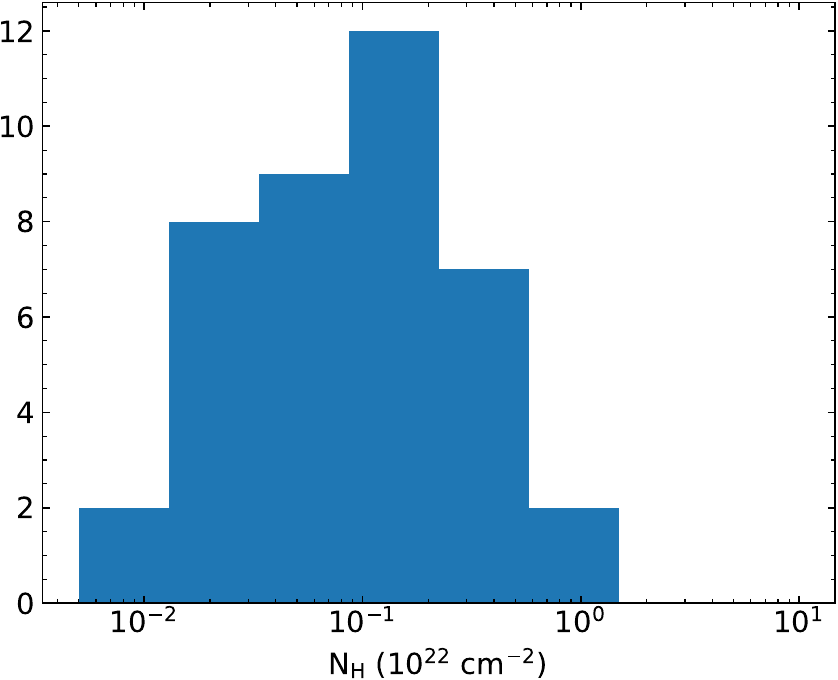}} \\ \vspace{0.3cm}
    \resizebox{.75\hsize}{!}
    {\includegraphics[]{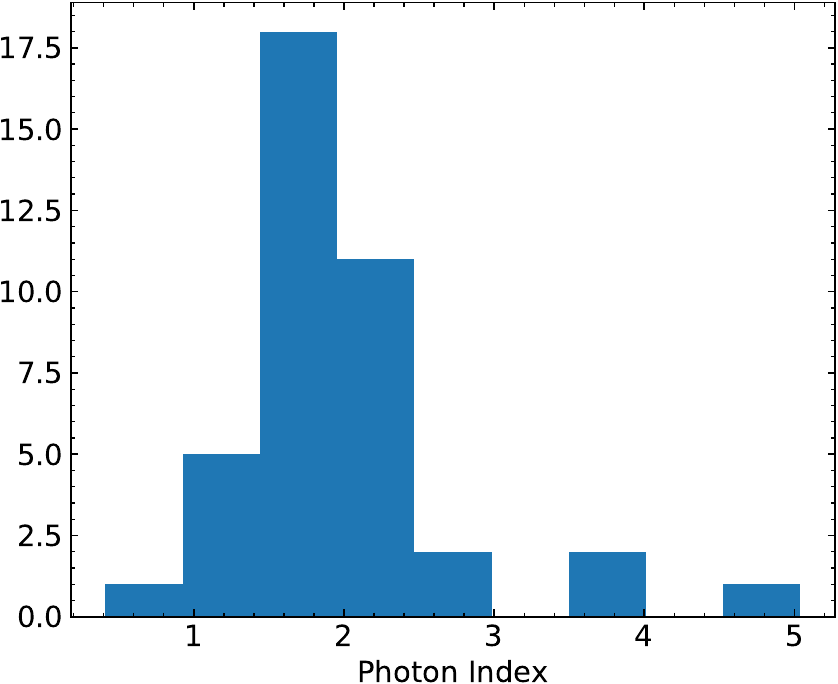}}
    \caption{Histograms of the distributions of the best-fit values of the absorption column densities (top panel) and photon indices (bottom panel) from the spectral fits of all the source on the HLD in Fig.\,\ref{fig:HLD}. 
    }
    \label{fig:histos}
\end{figure}

We computed unabsorbed fluxes in the 0.3--1.5\,keV (soft) and 1.5--10\,keV (hard) energy ranges together with the corresponding hard-to-soft ratio, as typically done for ULXs \citep[see e.g.,][]{Gurpide2021a_ULXsample}. From the broadband unabsorbed flux, compute in the energy range 0.3--10\,keV, we derived the unabsorbed luminosities, using the distances of the host galaxies of \citet{Tranin2024_cat}, derived from GLADE. The resulting HLD is shown in Fig.\,\ref{fig:HLD}, while Fig.\,\ref{fig:histos} displays the distribution of the absorption columns and photon indices. 
The median $N_\mathrm{H}$ is $1.1\times10^{21}$\,cm$^{-2}$, while the median photon index is 1.85. 
As a result of fitting with free parameters, the broadband fluxes for two sources turned out lower than those reported in the 4XMM catalogue and below the HLX threshold of $10^{41}$\,\ergs{}, leading to their reclassification as ULXs. Those sources can be considered extreme ULXs \citep[eULXs,][]{Gladstone2013} and are still of interest because they are typically hard and are considered good candidates for IMBHs.

\section{ESO 243-49 HLX-1}
\label{sec:HLX1}

\begin{table*}
      \caption[]{Log of archival \xmm{} observations of \hlxone{}. }
         \label{tab:LogObs}
     $$ 
         \begin{small}
         \begin{tabular}{lccccc}
         \hline\hline\noalign{\smallskip}
            ObsID. & Date & EPIC instr. & Exposure times (ks) & EPIC counts\\
            \noalign{\smallskip}
            \hline\hline
            \noalign{\smallskip}
            0204540201 & 2004-11-23 & MOS1/MOS2 & 21.6/21.6 & $2640\pm66$ \\
            \noalign{\smallskip}
            0560180901 & 2008-11-28 & pn/MOS1/MOS2 & 35.3/49.9/49.9 & $14698\pm130$\\
            \noalign{\smallskip}
            0655510201 & 2010-05-14 & pn/MOS1/MOS2 & 74.1/104.4/104.5 & $1185\pm48$\\
            \noalign{\smallskip}
            0693060401 & 2012-11-23 & pn/MOS1/MOS2 & 43.5/50.7/50.8 & $3385\pm66$\\
            \noalign{\smallskip}
            0693060301 & 2013-07-04 & pn/MOS1/MOS2 & 108.4/120.2/120.3 & $2469\pm63$\\
            \noalign{\smallskip}
            0801310101 & 2017-05-17 & pn/MOS1/MOS2 & 110.8/124.5/127.5 & $42910\pm221$\\
            \noalign{\smallskip}
            0844030101 & 2021-01-01 & pn & 17.7 & $121\pm16$\\
            \noalign{\smallskip}
            0891802101 & 2021-11-16 & -- & 14.8 & $58\pm12$\\
            \noalign{\smallskip}
            \hline\hline
            \end{tabular}
            \end{small}
        $$ 
        \tablefoot{The reported EPIC instruments are those where the source was detected and a spectrum was extracted. The exposure times correspond to the duration of the observation after the subtraction of the flaring background, except for ObsID.0891802101, where no spectrum was available. The EPIC counts refer to the estimated source counts in all EPIC detectors in the 0.2-12\,keV band. }
   \end{table*}   

\begin{table*}
      \caption[]{Best-fit values of \xmm{} spectra of \hlxone\ used in this work. }
         \label{tab:HLX-1Fit}
     $$ 
         \begin{tabular}{lcccccccc}
         \hline\hline\noalign{\smallskip}
            ObsID. & $N_\mathrm{H}$ & $T_\mathrm{disc}$ & $\Gamma$ & HR & Unabs. $L_\mathrm{X
            }$ & $\chi^2$/d.o.f.\tablefootmark{b} & n.h.p. & State\\
            & ($10^{20}$\,cm$^{-2}$) & (keV) &&& ($10^{41}$\,\ergs{}) && \\
            \noalign{\smallskip}
            \hline\hline
            \noalign{\smallskip}
            0204540201 & $15_{-10}^{+14}$ & $0.09_{-0.02}^{+0.12}$ & $3.4\pm0.6$ & $0.080\pm0.006$ & $7.3\pm0.4$ & 45.81/41 & 0.279 & High/soft \\          \noalign{\smallskip}
            0560180901 & 1.52\tablefootmark{a} & $0.134\pm0.003$ & $2.53_{-0.19}^{+0.18}$ & $0.160\pm0.004$ & $3.82\pm0.06$ & 144.62/150 & 0.609 & High/soft\\
            \noalign{\smallskip}
            0655510201 & 1.52\tablefootmark{a} & $0.20\pm0.02$ & $0.9_{-0.8}^{+0.7}$ & $1.48\pm0.18$ & $0.225\pm0.019$ & 79.74/60 & 0.045 & Low/hard\\
            \noalign{\smallskip}
            0693060401 & 1.52\tablefootmark{a} & $0.127_{-0.014}^{+0.013}$ & $2.08\pm0.19$ & $0.66\pm0.03$ & $0.91\pm0.03$ & 86.61/82 & 0.343 & Interm.\\
            \noalign{\smallskip}
            0693060301 & 1.52\tablefootmark{a} & $0.16_{-0.03}^{+0.04}$ & $2.0_{-0.3}^{+0.2}$ & $0.83\pm0.06$ & $0.284\pm0.015$ & 110.39/109 & 0.445 & Low/hard\\
            \noalign{\smallskip}
            0801310101 & $3.1\pm1.3$ & $0.142\pm0.004$ & $2.9\pm0.16$ & $0.1393\pm0.0017$ & $5.36\pm0.05$ & 334.74/281 & 0.015 & High/soft \\
            \noalign{\smallskip}
            \hline\hline
            \end{tabular}
        $$ 
        \tablefoot{The HR is the ratio of the unabsorbed fluxes in the energy bands 1.5--10\,keV and 0.3--1.5\,keV. Unabsorbed luminosities are computed in the 0.3--10\,keV band.
         \tablefoottext{a}{Parameter frozen to the Galactic absorption \citep{HI4PICollaboration2016}.}
         \tablefoottext{b}{d.o.f. $=$ degrees of freedom.}
         }
   \end{table*}

Compared to other HLXs, \hlxone{} has been well studied and observed in the past decades, with different missions, so that it is possible to study the evolution of this source on yearly timescales \citep[e.g.][]{Webb2023_TDEQPE,Lin2020_HLX1monitor}. In particular, many authors demonstrated that the spectral evolution of \hlxone{} is similar to that of an X-ray BH binary, with transitions from high/soft to low/hard states \citep{Godet2009,Servillat2011,Godet2012,Webb2012,Straub2014,Soria2021}. 
Hence, to achieve a more complete understanding of the behaviour of this source and looking for similarities within our sample of sources, we considered all archival \xmm{} observations of \hlxone{} (see Table \ref{tab:LogObs}; note that only ObsID.0801310101 is included in our sample, as it is the only observation present in the catalogue of \citealp{Tranin2024_cat}). 
Following the previous analysis (cf. Sect.\,\ref{sec:SpectralAnalysis}), we retrieved all EPIC spectra of \hlxone{} from the 4XMM-DR12 and fitted the spectra of each observation individually. ObsID.0204540201 has no pn spectrum because the source falls on a CCD gap. In ObsID.0844030101 the source was detected only in the EPIC-pn, most likely because of its faintness and a short exposure. In this observation, the spectrum has $\lesssim$100 counts in the energy range 0.3--10\,keV and the background dominates above 1\,keV, and therefore we decided to exclude it from the analysis. ObsID.089180201 had no spectrum available, implying that although the source was detected the spectrum had less than 100 counts. To our knowledge, this is the first time that the results of the fit of the 2017 data set are presented in a paper.

The simple power-law fit already showed the spectrum to be more complex, as the fit was statistically rejected ($\chi^2>2$). This is due to the presence of a soft excess at $\sim$0.7\,keV, typically modelled with an additional a thermal component \citep{Servillat2011,Godet2009}. We tried with a simple black body and with a multi-temperature disc black body  (\texttt{diskbb}). Whether or not these are the best models and what their implications are is  beyond the scope of this paper. The objective is to have a good enough description of all \hlxone{} spectral states, to allow for a comparison with all the HLX candidates in the HLD. Among the two, the model with the black body returned statistically good fits for all the spectra. 
Best-fit results are reported in Table \ref{tab:HLX-1Fit}. We also computed the HR as in Sect.\,\ref{sec:SpectralAnalysis} and the unabsorbed X-ray luminosities assuming a distance of the host galaxy of 90.94\,Mpc, as in \citet{Tranin2024_cat}. We plot the spectral evolution of \hlxone{} on the HLD in Fig.\,\ref{fig:HLD} (red squares and arrows). We also noted that the spectrum of \hlxone{} is significantly dominated by the background at energies $\gtrsim$2\,keV ($\gtrsim$5\,keV for ObsIDs.\,0560180901 and 0693060401). However, we decided to perform the fit in the usual energy band 0.3--10\,keV, to try to constrain the photon index of the power law and be consistent with the analysis of the sample (Section\,\ref{sec:SpectralAnalysis}).

\section{Analysis of selected candidates}
\label{sec:AnalysisSelectedSources}

The majority of our sample has HRs in the range 0.5--5, with a few outliers. Two sources stand out for their high HR values ($\geq10$): 4XMM J050252.9-755108 (ObsID.0762800101) and 4XMM J050117.1-242808 (ObsID.0110870101). Both source spectra have few spectral bins (4 and 5, respectively) in the energy range 0.3--10\,keV, so we caution the reader about the reliability of these data point. The \xmm{} position of the first one matches those of three \textit{Gaia} counterparts, one suggested to be a galaxy (with a probability of 1.0), possibly the host galaxy (distance of 367\,Mpc), the other two stars (probabilities $\geq0.99$) of magnitude $\sim$20 \citep{GaiaEDR2020}. The second source has a robust association with the galaxy Gaia DR3 2960239864411377536  \citep{GaiaEDR2020}, at the distance of 975\,Mpc. The lack of any additional X-ray or multiwavelength data prevent us from deducting more on the nature of these two sources. 

The softest sources in our sample are particularly interesting candidates. For each of them, we conducted a deeper analysis, presented in the following subsections, incorporating X-ray and multiwavelength data from other missions, when available. Information about the host galaxies was obtained from the SIMBAD Astronomical Database\footnote{\url{https://simbad.cds.unistra.fr/simbad/}.}.

\subsection{\candone}
\label{subsec:J2318}

This source is associated with the galaxy 2MASX J23181913-4222355 (at 260\,Mpc) and lies at 5\,kpc ($\sim$5") from the galaxy centre. The host has been spectrally identified as a passive galaxy 
by the 6dF DR3 survey \citep{Jones2009_6dFGalaxySurvey}, excluding the possibility of a low-luminosity AGN. The estimated unabsorbed flux of this source in the 0.3--10\,keV band is $(1.6_{-0.5}^{+0.7}) \times10^{-13}$\,\ergcms, which results in an unabsorbed luminosity  $L_\mathrm{X}=(1.3_{-0.4}^{+0.6})\times10^{42}$\,\ergs{} if associated with this galaxy. Hence, this candidate would classify as a bona-fide HLX. 

However, a visual inspection of the field of view of the source revealed that it also lays in the halo of a closer galaxy, NGC 7582, a barred spiral galaxy with a Seyfert 2 AGN at $\sim$17.6\,Mpc \citep{Liu2005}, 5.1\,kpc away from the centre. 
If hosted in NGC 7582, the estimated unabsorbed X-ray luminosity of the source would be of $6\times10^{39}$\,\ergs{}, in the ULX regime.

\begin{figure}
    \centering
    \includegraphics[width=0.45\textwidth]{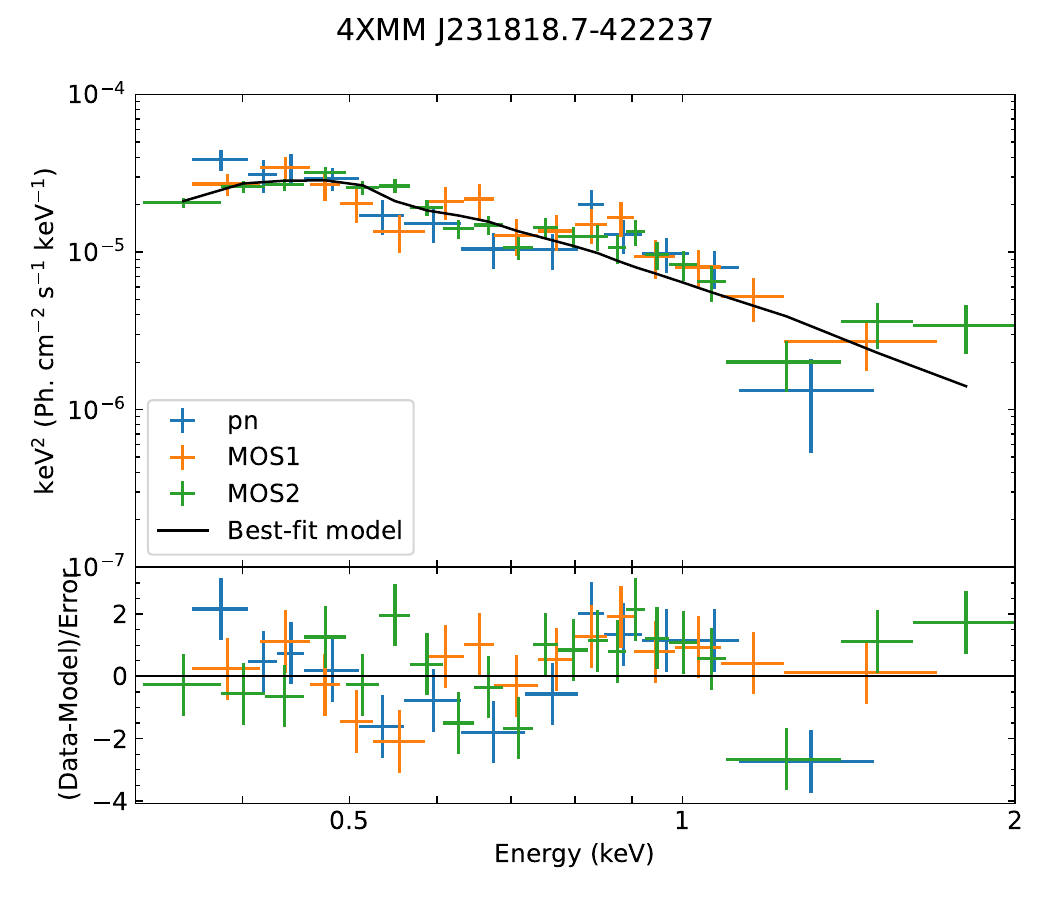}
    \caption{Fit of the EPIC spectrum of 4XMM J231818.7-422237. Top panel: pn (blue), MOS1 (orange) and MOS2 (green) data points and best-fit absorbed power law model (black line). Bottom panel: Residuals of the fit. The spectrum has been rebinned for a better visualisation. The background becomes dominant above $\gtrsim$1\,keV.}
    \label{fig:Fit_J2318}
\end{figure}

There are in total five \xmm{} archival observation of the field of view of \candone, taken in the years 2001, 2005, 2007, 2016, and 2018, but the source is detected only in the 2005 observation (ObsID.0204610101). 
The spectral fit of this source (Fig.\,\ref{fig:Fit_J2318}) resulted in $N_\mathrm{H}=(1.3\pm0.4)\times10^{21}$\,cm$^{-2}$ and $\Gamma=5.0\pm0.4$ ($\chi^2$/d.o.f.=157.1/122). With an HR of $(7.5\pm0.4)\times10^{-3}$, it is the softest source of the sample. The addition of a thermal component (a black body or a multi-temperature disc black body) to the power law model did not improve the fit. An absorbed single-thermal component model returned an unacceptable fit (n.h.p.$<$0.01). 

\begin{figure}
    \centering
    \includegraphics[width=\columnwidth]{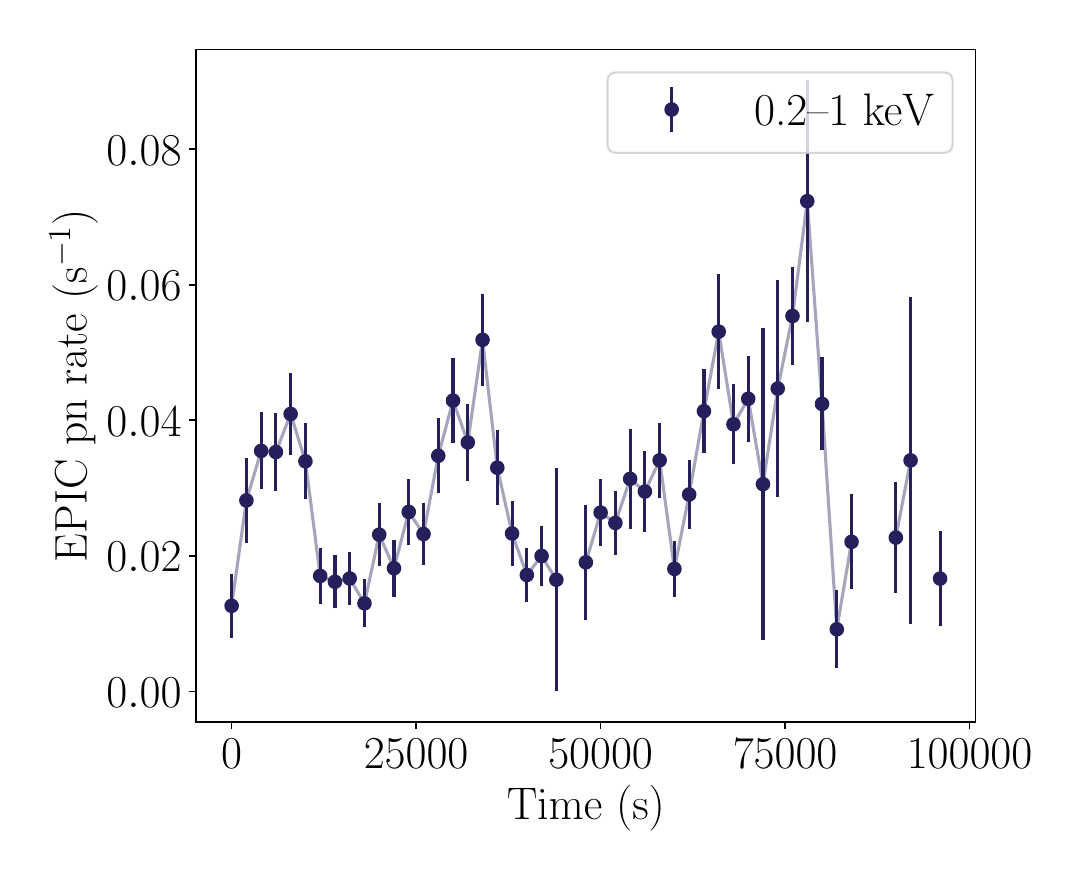}
    \caption{Short-term background subtracted EPIC pn 0.2--1\,keV lightcurve of 4XMM J231818.7-422237 (ObsID.020461010), binned at 2\,ks.}
    \label{fig:ShortTerm_J2318}
\end{figure}

The upper limits from the other observations, computed using the FLIX sensitivity estimator \citep{webb_xmm2athena_nodate}, lead to a long-term variability larger than by a factor of 10. The source is also variable on short timescales in its only detection, showing oscillations with an amplitude of a factor of $\sim$5 on a typical timescale of $\sim$20\,ks. For this reason, the source was noted of particular interest in \citet{lin_classification_2013}. The 0.2-1.0 keV EPIC pn light curve, obtained after standard filtering similar to Sec. \ref{sec:HLX1}, is shown in Fig. \ref{fig:ShortTerm_J2318}. The corresponding normalised excess variance \citep[computed as described in][]{Vaughan2003} is $0.11\pm0.03$. There is no significant spectral change linked to this variability. A search for pulsations from this source has already been performed in the framework of the EXTraS project \citep{DeLuca2021_extras} and no signal has been found. A more sophisticated search, accounting for instance for orbital or first period derivative corrections, is beyond the goals of this paper. Here, we computed the upper limit on the pulsed fraction (PF), defined as the modulation amplitude divided by the average of that modulation, for the EPIC-pn time series (the MOS cameras have lower temporal resolution), with the software package \textit{Stingray} \citep{Bachetti2022_stingray,Huppenkothen2019}. Assuming a sinusoidal pulsed signal  \citep[typical for ULXs, e.g.][]{Bachetti2014}, for a power of 40 and 1 harmonic, we obtained a PF$\lesssim40$\%, at a confidence level of 95\%.

\subsection{\candtwo}
\label{subsec:J1615}

\begin{figure}
    \centering
    \includegraphics[width=0.45\textwidth]{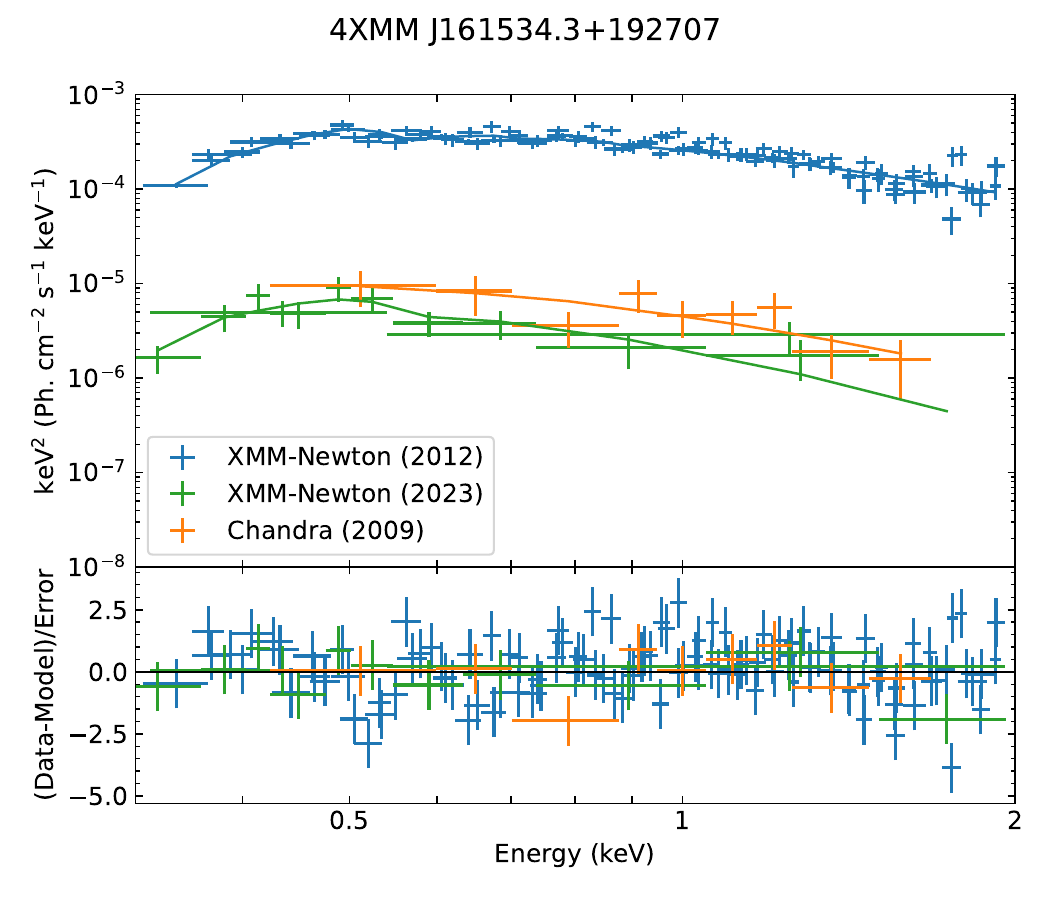}
    \caption{Best fit of the \xmm\ (2012 in blue and 2023 in green) and \chandra\ (2009 in orange) spectra of \candtwo\ (top panel) and residuals of the fits (bottom panel). The 2023 \xmm\ and \chandra\ spectra were rebinned for a better visualisation.}
    \label{fig:Fit_J1615}
\end{figure}

This source is associated with the radio galaxy NGC 6099, at 133\,Mpc (the source lies 17.54$^{\prime\prime}$ from the galaxy centre), in the galaxy cluster 1RXS J161531.4$+$192754. 
It is also present in the CSC2 catalogue, thanks to a \chandra{} observation taken in 2009 and centred on NGC 6099. It was already classified as an extreme luminous X-ray source candidate in the catalogue of \citet{Gong2016}, with an absorbed luminosity of $\sim3\times10^{40}$\,\ergs{}, in the energy range 0.3--8\,keV. Recently, it was reported as an IMBH candidate by \citet{Chang2025}. The field of view of \candtwo\ was observed twice with \xmm\, in 2012 (ObsID.0670350301) and 2023 (ObsID.0923030101). 
The source was detected in the 2012 observation and its spectrum is present in our sample. The spectral fit results in an unabsorbed X-ray luminosity $L_\mathrm{X}=(6.7_{-0.7}^{+0.9})\times10^{42}$\,\ergs{} (0.3--10\,keV range). The best-fit parameters are $N_\mathrm{H}=(1.86_{-0.17}^{+0.18})\times10^{21}$\,cm$^{-2}$ and $\Gamma=4.0_{-0.11}^{+0.12}$, with an HR of $(4.11\pm0.11)\times10^{-2}$ ($\chi^2$/d.o.f.=177.6/141, n.h.p.$=$0.02). The best-fit is  shown in Fig.\,\ref{fig:Fit_J1615}. The use of a thermal component instead of or in addition to the power law did not significantly improve the fit statistics. The resulting unabsorbed luminosity is 2--3 times higher than that estimated by \citet{Chang2025}, who used different fitting models. As for the previous candidate, the EPIC-pn observation did not show pulsations and we estimated an upper limit on the PF of $\sim$25\%.

In the 2023 observation the source is not detected by the \xmm\ source detection pipeline, likely because of a crowded field within diffuse emission, but it is clearly visible by eye in the EPIC images. Hence, we extracted manually the source spectrum. First, we reprocessed the non-calibrated data with the Science Analysis System (SAS), version 21.0.0, and Current Calibration Files (CCF) released in November 2023. For the EPIC-pn and MOS1 images, we extracted the source spectrum from a circular region centred on the source coordinates, with a radius of 8$^{\prime\prime}$ to prevent contamination from the the diffuse emission of the host galaxy. The background spectrum was extracted from a circular region of $\sim20^{\prime\prime}$ enclosing the X-ray nebular region surrounding the source, with the subtraction of two circular regions for the HLX candidate and the host galaxy. The source is not visible in the MOS2 data, where the diffuse emission dominates. We rebinned the pn and MOS1 spectra with a minimum of 20 counts per bin and fitted them together with the usual absorbed power law model, in the energy range 0.3--2\,keV (the X-ray background dominates at energies $\gtrsim2$\,keV). The lower bound of the absorption was unconstrained and we fixed the $n_\mathrm{H}$ to the best-fit value of the 2012 \xmm\ observation. We obtained a photon index $\Gamma=5.5_{-0.7}^{+0.9}$, with a $\chi^2$/d.o.f.$=$24.26/23. The unabsorbed flux is $F_\mathrm{X}=(6.0\pm1.0)\times10^{-14}$\,\ergcms{}, corresponding to $L_\mathrm{X}=(1.3\pm0.2)\times10^{41}$ \ergs, in the 0.3--10\,keV range. The corresponding HR is $(3.4\pm0.8)\times10^{-3}$. The 2023 detection is not in the EXTraS archive. A preliminary inspection of the power density spectrum revealed no significant peaks, which is expected given the source's low count statistics ($\sim$500 counts). The upper limit on the PF was $\sim$70\%.

For the \chandra\ observation of 2009, first we reprocessed the data with the \texttt{chandra\_repro} script of the \chandra{} data analysis software CIAO, v.\,4.15.1, with CALDB v.\,4.10.4 \citep{Fruscione2006}. We extracted the source and background spectra with the CIAO script \texttt{specextract}, selecting a circular region of radius 0.8$^{\prime\prime}$ centred on the source coordinates for the source spectrum and a circular region of radius 10$^{\prime\prime}$ on the same CCD in an area free of other X-ray sources for the background spectrum. We fitted the background subtracted spectrum with the same absorbed power law model, using Cash statistics \citep{Cash1979}, since the source has 56 net counts in the 0.5--8\,keV range. The $N_\mathrm{H}$ was unconstrained and we fixed it to the best-fit value of the 2012 \xmm\ observation. We obtained a photon index of $\Gamma=4.5\pm0.8$ and an unabsorbed flux of $F_\mathrm{X}=(7.8\pm1.9)\times10^{-14}$\,\ergcms{} (C-stat/d.o.f.$=$31.51/44), corresponding to $L_\mathrm{X}=(1.7\pm0.4)\times10^{41}$ \ergs, in the 0.3--10\,keV range, and consistent with that of the 2023 \xmm\ observation. The corresponding HR is $(1.8\pm0.6)\times10^{-2}$. The best-fit values of the 2009 and 2023 spectra, where the source was caught in a dimmer state (see Fig.\,\ref{fig:Fit_J1615}), are consistent or marginally consistent with the results of \citet{Chang2025}. \candtwo\ displays a clear long-term variability (see Fig.\,\ref{fig:HLD}) and it is notably soft, suggesting similarities with \hlxone{}.

\begin{figure}
    \centering
    \includegraphics[width=1\linewidth]{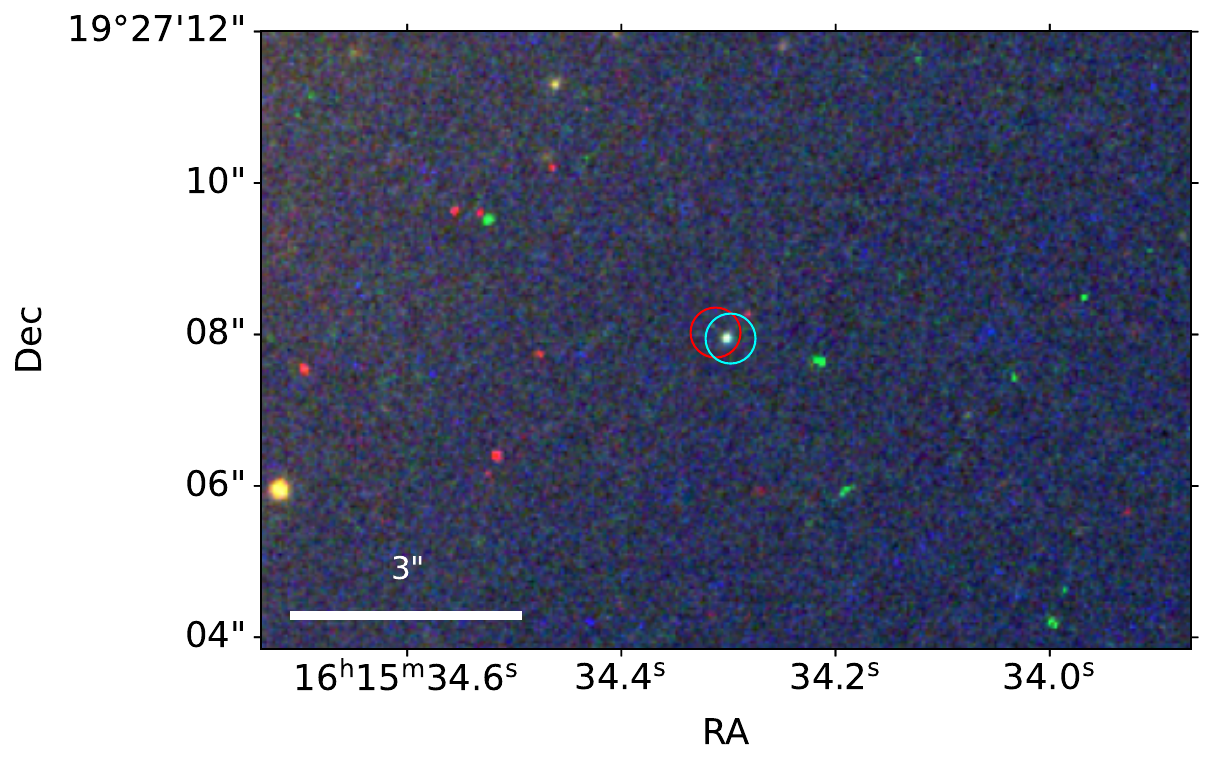}
    \caption{HST cutout centred on the position of \candtwo. The colour composite is based on HST observations in the F300X, F475W, and F814W filters taken in September 2023. The error circles represent the position and uncertainty reported in the \chandra{} (cyan) and \xmm{} (red) catalogues.}
    \label{fig:hst_cutout}
\end{figure}

Further supporting this similarity, we identified an optical counterpart in an archival \textit{Hubble Space Telescope} (HST) exposure from September 2023 (Proposal 17177, PI: I. Chilingarian).  
Fig.\ref{fig:hst_cutout} presents a colour composite of HST observations, revealing a point-like source located precisely at the intersection of the \xmm{} and \chandra{} error circles. The source is also tentatively visible in Legacy Survey DR9 images, with consistent magnitudes in both datasets: $m_{F300X} = 26.1 \pm 0.5$, $m_{F475W} = 24.8 \pm 0.4$ and $m_{F814W} = 24.3 \pm 0.4$. Assuming the source is an unresolved star cluster associated with NGC 6099 (its high optical luminosity prevents it from being a single star), with a mass-to-light ratio of $\sim$2 M$_\sun$/L$_\sun$ and a Galactic extinction of $A_V = 0.13$, its estimated mass would be $\sim 6 \times 10^6$ M$_\sun$. Alternatively, the optical emission can originate from an irradiated accretion disc \citep[see][for a detailed discussion on both possibilities]{Chang2025}. 

\subsection{\candthree}
\label{subsec:candthree}

This source is associated with the galaxy Gaia DR3 1323128966899690752, $\sim$570\,Mpc from Earth. There is no counterpart in any other wavelength. It has been detected with \xmm{} 6 times, from 2003 to 2016, but a spectrum is only available for one observation, taken in 2015 (ObsID.0764460201). The fit returned the best values of $\Gamma=3.5\pm0.4$, with a fixed $N_\mathrm{H}=2.5\times10^{20}$\,cm$^{-2}$ ($\chi^2$/d.o.f.=19.24/12, n.h.p.$=$0.08), an unabsorbed flux of $(9.2\pm1.2)\times10^{-15}$\,\ergcms, corresponding to a luminosity  $L_\mathrm{X}=(3.6\pm0.5)\times10^{41}$\,\ergs. The corresponding HR was $(8.8\pm1.6)\times10^{-2}$ and it falls among the softest sources, very close on the HLD to the position of \hlxone\ when in the high/soft state. No pulsation is detected for this source and the PF upper limit is $\sim$75\%. The six \xmm\ detections report an absorbed flux as computed in the 4XMM catalogue between $3.3\times10^{-15}$\,\ergcms and $1.3\times10^{-14}$\,\ergcms. With no other information on its spectral variability it is hard to classify the source, but its luminosity and hardness would suggest a similar nature to \hlxone{}.

\subsection{\candfour}
\label{subsec:candfour}
\candfour\ is a relatively hard source located in the outskirts of the galaxy ESO 516-9, at a distance of $\sim110$\,Mpc. 
It has already been identified as a ULX/HLX candidate thanks to several catalogue studies \citep{Earnshaw2019,Bernadich2022,Tranin2024_cat}. There are two \xmm{} observations of this object, two days apart in 2008, that do not show any significant variability; \erosita\ also provides a more recent unconstraining upper limit of $1.61\times\,10^{-13}$\,\ergcms\ in the 0.2-5 keV band. The \xmm{} spectrum is consistent with a power law with $\Gamma=2.3_{-0.6}^{+0.8}$ ($N_\mathrm{H}=(1.9_{-1.4}^{+1.8})\times10^{21}$\,cm$^{-2}$, $\chi^2$/d.o.f.=32.36/21, n.h.p.$=$0.05), with a resulting unabsorbed flux of $(2.3_{-0.5}^{+1.7})\times10^{-14}$\,\ergcms and a corresponding luminosity $L_\mathrm{X}=(3.4_{-0.7}^{+2.5})\times10^{40}$\,\ergs\ in the 0.3--10 keV range, if it is indeed associated with ESO 516-9. In spite of being below the conventional luminosity threshold for HLXs, this source occupies the same HLD space as \hlxone\ when in the low/hard state, suggesting yet another source of the same type. Also this source does not display any pulsating signal, with a PF upper limit of $\sim$80\% (the source has less than 250 counts).

There is no clear multi-wavelength counterpart to the X-ray source within the extent of ESO 516-9. The only apparent possible counterpart would be the infrared source WISE J161603.96-223726, which is detected as separate from the galaxy's extended emission by the automatic pipeline (notably through its significantly softer emission, with a brighter emission in the bands 3 and 4 of WISE) -- although its separation from the galaxy's emission is not visually clear. This object could thus possibly be a background AGN behind the foreground galaxy. Spatially resolved optical spectra (e.g. through an integral field unit) are needed to assess for instance the presence of emission at a higher redshift around the location of the X-ray source (which would indicate it is a background AGN).

\section{Discussion}
\label{sec:Discussion}

We have investigated the population of HLXs observed by \xmm\ and collected in \cite{Tranin2024_cat} to determine whether HLXs are just ``more extreme'' ULXs, and as such, whether they share the same properties of ULXs in terms of spectral shapes and evolution, or if they are a different category of astrophysical sources with different properties. 
After retrieving the spectra of individual sources and removing those securely identified as contaminants (mostly galaxy pairs), we fitted the remaining ones with an absorbed power law, to build the HLD. The final sample consists of 40 individual sources. As expected, the fluxes and HRs that we obtained are slightly different from those in the 4XMM-DR12 catalogue, since we left the fit parameters free to vary (cf.\,Sect.\,\ref{sec:SpectralAnalysis}). This impacts also the luminosities of the selected sources, which are slightly different than those reported in \citet{Tranin2024_cat}.

The space on the HLD occupied by the HLX sample overlaps with the space where a typical AGN with $\Gamma=2$ would fall (vertical red area in Fig.\,\ref{fig:HLD}, cf.\,Sect.\,\ref{sec:SpectralAnalysis}), highlighting the intrinsic difficulties in disentangling the two classes of objects. 
As a results, some AGNs can contaminate ULX and HLX catalogues \citep{Tranin2024_cat, Walton2022_ULXcat, Kovlakas2020_ULXcat}. This issue becomes more problematic at greater distances, due to increasing difficulties in resolving objects close to each other. Typically, HLXs are detected at greater distances than ULXs, increasing the chances of ambiguous associations. 
From our study, we can confirm that 10 out of 64 initial spectra belong to AGNs in galaxy pairs wrongly associated with the companion galaxy, unresolved groups of galaxies, or intra-galactic diffused emission. 
In total, 25\% of the retrieved spectra were of sources labelled as ``galaxy pairs'' in the catalogue of \citet{Tranin2024_cat} and were excluded from the analysis presented here. These spectra significantly differ from the spectra of the final sample in shape and number of counts. In fact, AGNs typically reach luminosities higher than $10^{41}$ \ergs{} and are the targets of long observations, so that their spectra have higher statistics, with a spectral shape that significantly deviates from a simple power law. It cannot be excluded that other AGNs remain in the final sample, especially if their spectra have low statistics. Nonetheless, the sample presented here can be considered one of cleanest produced to date. 

The bulk of the sample consists of sources with HRs in the range 0.5--5 and luminosities between $\sim10^{41}$\,\ergs{} and $10^{43}$\,\ergs, with no particular distribution between sources with ``robust'' or ``weak'' associations to their host galaxies as in the original catalogue. Approximately 50\% of sources are harder than typical AGNs, hinting to the possibility that half the sample might actually be an extension of the ULX population towards higher luminosity. Indeed, when fitted with the same model, ULX spectra are typically harder than the HLX sample analysed in this work, with $\Gamma<2.3$ \citep{Gladstone2009}. On the other hand, roughly 40\% of the HLX candidates have spectra with HR$\leq$1, indicative of a significant soft part of the population. These candidates could represent a distinct class of objects, rather than typical ULXs, that encompasses astrophysical phenomena reaching hyperluminous levels, such as TDEs.

A prime example would be \hlxone{}, present in the catalogue, interpreted by many authors as an IMBH, with mass estimations in the range $10^{4}$--$10^{5}$\,M$_\odot$ \citep[e.g.][]{Webb2012}, undergoing a repetitive partial TDE \citep{Godet2014,Webb2014}. There is also one other source in the sample, 4XMM J215022.4-055109, interpreted as a TDE around an IMBH in an off-centre star cluster \citep{Lin2018}. This suggests that other TDE candidates can hide among the softest HLXs. Indeed, we found four more candidates showing properties similar to \hlxone{}. The most striking is \candtwo\ (cf. Sect.\,\ref{subsec:J1615}), the most luminous and the second softest source of the sample. It occupies an area on the HLD close to \hlxone{} in the high/soft state and displays a clear long-term variability (see Fig.\,\ref{fig:HLD}), with a difference in luminosity of almost two orders of magnitudes. It is associated to an optical counterpart, which can be interpreted as a star cluster. Therefore, the interpretation of \candtwo\ as a partial TDE, similar to \hlxone{}, is plausible \citep[as also concluded in the recent work of][]{Chang2025}. However, the evolution of the HR of \candtwo\ differs from that of \hlxone{} and is more akin to the behaviour of hard ULXs, which transition from a hard, high-luminosity state to a softer, dimmer state \citep[see the examples in][]{Gurpide2021a_ULXsample}.

Two other soft candidates, \candthree\ (Sect.\,\ref{subsec:candthree}) and \candfour\ (Sect.\,\ref{subsec:candfour}) , occupy the same position in the HLD  of \hlxone\ in the high/soft and low/hard states, respectively. The available flux estimates of \candthree\ do not suggest changes in luminosities by orders of magnitude, but this might be due to the limited coverage of the source. For both sources, the lack of more high-quality data does not allow for a more secure classification. Lastly, the softest source of the sample, \candone\  is an ambiguous case, because it can be yet another soft HLX or a SSUL sources, based on its possible association with its host galaxy (cf. Sect.\,\ref{subsec:J2318}). The pronounced variability in the source light curve is reminiscent of the temporal variability of other TDEs. Unfortunately, the lack of a long-term coverage of the source prevents us to advance more secure claims, as done for example for the TDE Swift J1644+57, which alternates between normal and dipping states, suggestive of a jet precession scenario \citep{Jin2021}.

None of the selected candidates show pulsations, which should not be surprising, considering that known PULXs have on average harder spectra \citep[see e.g. the HLD in][]{Gurpide2021a_ULXsample}, compared to the candidates analysed here. Furthermore, the limited source statistics and the transient nature of the pulsed signals hinder the detection of pulsations. Notably, the prototype of this class, \hlxone, does not exhibit pulsations despite having higher count statistics. Taken together, these considerations suggest that very soft HLXs may constitute a distinct class of objects, differing from ``standard'' ULXs and possibly representing IMBH and/or TDEs.

The work presented in this paper is subject to a few biases and limitations. First, the sample is limited to those sources that have average HLX luminosity levels during an observation (cf.Sect.\,\ref{sec:SampleSelection}). Hence, sources that occasionally surpass the $10^{41}$\,\ergs{} luminosity threshold for a shorter time segment  than the duration of the observation are automatically excluded. Second, the sample is comprehensive of \xmm{} data alone. However, considering the lower-quality of the spectra collected by the other two X-ray missions, \swift/XRT and \chandra{}, and considering the lower number of their HLX candidates (88 vs. 115 of \xmm{}), the inclusion of their spectra in the present analysis probably would have not lead to substantial changes in the final results.

\section{Conclusions}
\label{sec:Conclusions}

This work investigates the spectral properties of a clean sample of HLXs as observed with \xmm{} and aims at discerning the presence of different type of sources within the same class. By performing spectral fits and correcting for the absorption along the line of sight, this is the first time that the intrinsic (i.e. unabsorbed) fluxes and luminosities of these objects are studied. The final sample is also the cleanest sample of \xmm\ HLXs produced to date, since the AGN contaminants have been identified and removed, not only on the basis of their association with the host galaxies, but also on their spectral properties. The resulting HLX population has luminosities in the range $10^{41}-10^{43}$\,\ergs\ and HR between 0.5 and 5, and it is overall softer than the population of ULXs, with no striking evidence for sub-populations. The only exceptions are a few very soft sources, which have similar spectral and temporal properties to \hlxone{}, suggesting that they differ in nature from ULXs and instead might resemble other type of sources, such as partial TDEs.

\begin{acknowledgements}
This research has made use of data obtained from the \textit{Chandra} Data Archive and the \textit{Chandra} Source Catalogue, and software provided by the \textit{Chandra} X-ray Center (CXC) in the application package CIAO; of observations obtained with \textit{XMM-Newton}, an ESA science mission with instruments and contributions directly funded by ESA Member States and NASA; of PyXspec, a Python interface to the XSPEC spectral-fitting program. The authors thank M. Coriat for hi assistance in providing the data. RA and GLI acknowledge financial support from INAF through grant ``INAF-Astronomy Fellowships in Italy 2022 - (GOG)''. 
\end{acknowledgements}

%
\bibliographystyle{aa} 
\bibliography{biblio} 
%

\begin{appendix}
\onecolumn

\section{Excluded sources}
\label{sec:ExcludedSources}

We collect here a few notes on all the spectra that were excluded from the final dataset because of a bad fit, determined by either a $\chi^2_\mathrm{red}>2$, or n.h.p.$<0.01$, or both. In a few cases, more complex models than a power law are needed to correctly fit the data, as for the high-count spectra of a few misclassified AGN present in the catalogue. In other cases, even with a low number of counts, the spectral shape significantly differ from a power law. Those spectra typically belong to spurious detections (e.g. intra-galaxy X-ray emission, detected as point-like sources) and hence are excluded from the sample. 
We recall that all spectra have at least 100 counts and are binned to have at least 5 bins of 20 counts each, suited to be fit with Gaussian statistics.

\paragraph{4XMM J114630.0+202728 (ObsID.0301900601)}. The automated fit resulted in a $\chi^2_\mathrm{red}>$2. We then manually fit the data and obtained an acceptable $\chi^2_\mathrm{red}$, but with a extremely high and unconstrained photon index. We try also to fix the $N_\mathrm{H}$ value to the Galactic one, but obtained again a $\chi^2_\mathrm{red}>$2. This source is hence excluded from of our sample. The source position matches that of the HII galaxy NAME Z 127-051 N, so the X-ray emission is probably due to the galaxy itself.

\paragraph{4XMM J163818.7-642201 and 4XMM J062648.7-543208 (ObsIDs.0093620101 and 0400010301)} These sources lie within significant X-ray diffuse emission, resulting in an intrinsic difficulties in separating the contribution of the diffuse emission to the source spectrum. 4XMM J163818.7-642201 belongs to a group of galaxies (ref. galaxy ESO 101--4). 4XMM J062648.7-543208 is detected both in the 4XMM-DR12 and the CSC2 catalogues and it is associated with the galaxy LEDA 19057, though at radial distance of $\sim$28 arcsec. Both sources are excluded from the sample.

\paragraph{4XMM J215022.4-055109 (ObsID.0823360101)} The fit resulted in a n.h.p. of only 0.01 and the source was thus excluded from our sample. Moreover, the source is already known in the literature as a TDE candidate \citep{Lin2018}, hinting to the chance that other TDE candidates might hide among HLXs.

\paragraph{4XMM J221755.3-354629 (ObsIDs.0201902901)} The automated fit resulted in a high reduced chi-squared and even fixing the $N_\mathrm{H}$ to the HI4PI map value did not improve it. Hence, we removed this source from the dataset. Overall the source is very soft and it lies in the outskirts of the galaxy LEDA 644746.

\paragraph{4XMM J011028.2-460422 (ObsID.0801310101)} This is \hlxone{} \citep{Farrell2009}. The automated fit resulted in $\chi^2_\mathrm{red}$(d.o.f.)$=1.959$(132) and a n.h.p. of the order of $10^{-10}$ (see also Table \ref{tab:LogObs}). In this observation the source is in soft state, in the HLX regime. More details about this source are given in Sect.\ref{sec:HLX1}.

\clearpage

\section{Data set}
\label{sec:Dataset}

We report in Table\,\ref{tab:TotalFit} a subset of all the data entries used in this work, with the X-ray luminosities and HRs obtained from the fit. The complete sample is available upon request.

\begin{table*}[h!]
\caption{List of HLX candidates analysed in this work.}
\label{tab:TotalFit}
\begin{center}
\begin{small}
\begin{tabular}{lccccccc}
\hline\hline
\noalign{\smallskip}
& ObsID & IAUNAME & Exposure Times & HR & X-ray Luminosity & Distance & Subsample \\
& & & (ks) & & ($10^{41}$\,\ergs{}) & (Mpc) & \\
\noalign{\smallskip}
\hline
\noalign{\smallskip}
1 & 0744410101 & 4XMM J192133.3+441322 & $26.7/34.3/34.3$ & $4.3\pm0.5$ & $30_{-3}^{+4}$ & 357 & robust \\
2 & 0044350101 & 4XMM J003416.6-212430 & $18.0/-/-$ & $4.54\pm1.5$ & $11_{-5}^{+7}$ & 269 & weak \\
3 & 0651460101 & 4XMM J210536.0-522300 & $37.3/-/-$ & $1.38\pm0.3$ & $0.6\pm0.2$ & 154 & robust \\
4 & 0555970401 & 4XMM J145753.6-113959 & $63.0/-73.6$ & $0.8\pm0.2$ & $4.6_{-1.4}^{+1.8}$ & 486 & weak \\
5 & 0306630201 & 4XMM J123115.8+105320 & $84.0/-/98.1$ & $1.\pm0.2$ & $7.8_{-1.3}^{+1.4}$ & 488 & weak \\
6 & 0204610101 & 4XMM J231818.7-422237 & $80./87.9/88.0$ & $0.0075\pm0.0004$ & $13_{-4}^{+6}$ & 260 & robust \\
7 & 0093550401 & 4XMM J233342.0-151523 & $17.8/21.3/22.2$ & $1.08\pm0.11$ & $8.1_{-0.8}^{+0.9}$ & 289 & weak \\
8 & 0827010901 & 4XMM J225931.7-344113 & $32.2/43.7/42.2$ & $0.8\pm0.09$ & $9.4_{-1.2}^{+2.2}$ & 263 & robust \\
9 & 0724690401 & 4XMM J015728.3-614120 & $10.7/-/-$ & $2.6\pm0.7$ & $1.0_{-0.3}^{+0.4}$ & 88 & robust \\
10 & 0743650801 & 4XMM J025645.2+413508 & $23.2/-/-$ & $2.2\pm0.5$ & $6.3_{-1.3}^{+3.2}$ & 281 & robust \\
11 & 0670350301 & 4XMM J161534.3+192707 & $13.0/16.1/16.1$ & $0.041\pm0.001$ & $67_{-7}^{+9}$ & 133 & robust \\
12 & 0744390301 & 4XMM J215057.1-073535 & $19.6/-/-$ & $2.3\pm0.8$ & $13.1_{-5.7}^{+1.3}$ & 617 & weak \\
13 & 0109910101 & 4XMM J140106.8-111128 & $39.2/-/-$ & $1.5\pm0.5$ & $3.14_{-1.08}^{+1.87}$ & 379 & robust \\
14 & 0675470501 & 4XMM J221014.5-121119 & $13.2/-/-$ & $2.85\pm1.02$ & $12_{-4}^{+5}$ & 403 & weak \\
15 & 0502630201 & 4XMM J085403.2+200746 & $49.6/54.7/54.8$ & $1.12\pm0.15$ & $1.7_{-0.3}^{+0.4}$ & 132 & weak \\
16 & 0505210401 & 4XMM J062536.4-535105 & $19.4/-/-$ & $2.1\pm0.8$ & $3.9_{-1.6}^{+2.6}$ & 247 & robust \\
17 & 0762800101 & 4XMM J050252.9-755108 & $11.7/-/-$ & $21\pm7$ & $49_{-23}^{+38}$ & 367 & weak \\
18 & 0800761301 & 4XMM J062447.9-372122 & $17.2/-/21.2$ & $1.5\pm0.2$ & $1.8\pm0.3$ & 139 & robust \\
19 & 0404050101 & 4XMM J074735.9+554115 & $96.3/-/-$ & $2.3\pm0.5$ & $7.2_{-1.5}^{+2.1}$ & 152 & weak \\
20 & 0203391001 & 4XMM J130307.6-154225 & $13.9/16.8/16.8$ & $2.0\pm0.4$ & $27_{-5}^{+6}$ & 324 & robust \\
21 & 0803910101a & 4XMM J161342.6+470314 & $79.6/-/-$ & $2.9\pm1.2$ & $3.2_{-1.3}^{+3.6}$ & 400 & robust \\
22 & 0604310401 & 4XMM J120335.1-040648 & $25.1/-/30.3$ & $2.8\pm0.2$ & $27\pm2$ & 287 & weak \\
23 & 0723100401 & 4XMM J125256.2-132756 & $60.5/-/-$ & $3.8\pm1.3$ & $33_{-14}^{+20}$ & 595 & robust \\
24 & 0552670301 & 4XMM J055239.0-363936 & $83.9/-/-$ & $0.46\pm0.16$ & $3.4_{-1.4}^{+2.5}$ & 200 & robust \\
25 & 0300530101 & 4XMM J125917.9+281107 & $-/-/25.0$ & $1.22\pm1.09$ & $5_{-4}^{+8}$ & 405 & weak \\
26 & 0764460201 & 4XMM J160553.3+324734 & $105.3/-/-$ & $0.088\pm0.016$ & $3.6\pm0.5$ & 573 & robust \\
27 & 0203170301 & 4XMM J123307.1+000758 & $77.9/102.9/103.2$ & $0.48\pm0.08$ & $17_{-4}^{+1}$ & 617 & robust \\
28 & 0782330301 & 4XMM J110106.5+102943 & $92.3/104.4/105.4$ & $2.06\pm0.12$ & $3.19\pm0.17$ & 171 & weak \\
29 & 0305980301 & 4XMM J042058.6-484008 & $10.9/15.5/14.8$ & $0.76\pm0.08$ & $58_{-7}^{+10}$ & 546 & weak \\
30 & 0555650301 & 4XMM J161604.0-223726 & $89.1/98.7/103.6$ & $0.65\pm0.11$ & $0.34_{-0.07}^{+0.25}$ & 111 & robust \\
31 & 0691590101b & 4XMM J222324.0-290444 & $54.7/-/-$ & $1.4\pm0.6$ & $1.0_{-0.5}^{+0.8}$ & 293 & robust \\
32 & 0107260101 & 4XMM J132811.0-313503 & $35.2/-/-$ & $2.24\pm1.12$ & $1.5_{-0.7}^{+0.9}$ & 184 & weak \\
33 & 0404410101 & 4XMM J011459.1+002519 & $37.5/-/49.8$ & $1.1\pm0.2$ & $1.3_{-0.3}^{+0.5}$ & 198 & robust \\
34 & 0083151201 & 4XMM J050107.9-384029 & $86.6/-/-$ & $1.9\pm0.6$ & $5.8_{-1.9}^{+2.5}$ & 288 & weak \\
35 & 0500940201 & 4XMM J101040.4+534719 & $23.6/-/-$ & $0.7\pm0.3$ & $20_{-8}^{+12}$ & 643 & robust \\
36 & 0149160201 & 4XMM J083235.2-225804 & $25.1/27.9/29.3$ & $0.7\pm0.1$ & $1.7_{-0.3}^{+0.7}$ & 118 & weak \\
37 & 0784610101 & 4XMM J194744.7+704645 & $21.8/-/-$ & $0.8\pm0.3$ & $24{-9}^{+14}$ & 878 & robust \\
38 & 0803910101b & 4XMM J161347.9+470809 & $78.9/-/-$ & $3.37\pm1.15$ & $1.4_{-0.5}^{+0.6}$ & 322 & robust \\
39 & 0201901801 & 4XMM J130256.1-022930 & $17.7/-/-$ & $1.6_{-0.5}^{+0.3}$ & $7.3_{-1.4}^{1.5}$ & 378 & weak \\
40 & 0110870101 & 4XMM J050117.1-242808 & $20.3/-/24.7$ & $7_{-4}^{+30}$ & $87_{-35}^{+232}$ & 1040 & robust \\
\noalign{\smallskip}
\hline\hline
\end{tabular}
\tablefoot{For each candidate we indicate the \xmm{} observation ID, IAU designation, exposure times for the EPIC-pn/MOS1/MOS2, hardness ratio, X-ray luminosity, host galaxy distance, and 'robust' or 'weak' classification label.}
\end{small}
\end{center}
\end{table*}

\end{appendix}

\end{document}